\documentclass[aip,pop,amsmath,amssymb,amsfonts,preprint
]{revtex4-1}

\usepackage{graphicx}
\usepackage{dcolumn}
\usepackage{bm}

\begin{document}

\preprint{AIP}

\title{Symmetries and Local Conservation Laws of Variational Schemes for the Surface Plasmon Polaritons}

\author{Qiang~Chen}
\email{cq0405@ustc.edu.cn}
\affiliation{State Key Laboratory of Complex Electromagnetic Environment Effects on Electronics 
and Information System, Luoyang, Henan 471000, China}
\affiliation{University of Science and Technology of China, Hefei, Anhui 230026, China}
\author{Xiaojun~Hao}
\affiliation{State Key Laboratory of Complex Electromagnetic Environment Effects on Electronics 
and Information System, Luoyang, Henan 471000, China}
\author{Chuanchuan~Wang}
\affiliation{State Key Laboratory of Complex Electromagnetic Environment Effects on Electronics 
and Information System, Luoyang, Henan 471000, China}
\author{Xiaoyang~Wang}
\affiliation{State Key Laboratory of Complex Electromagnetic Environment Effects on Electronics 
and Information System, Luoyang, Henan 471000, China}
\author{Xiang~Chen}
\affiliation{State Key Laboratory of Complex Electromagnetic Environment Effects on Electronics 
and Information System, Luoyang, Henan 471000, China}
\author{Lifei~Geng}
\affiliation{State Key Laboratory of Complex Electromagnetic Environment Effects on Electronics 
and Information System, Luoyang, Henan 471000, China}

\date{\today}

\begin{abstract}
The relation between symmetry and local conservation law, known as Noether's theorem, plays an important role in 
modern theoretical physics. As a discrete analog of the differentiable physical system, a good numerical scheme 
should admit the discrete local conservation laws and inherent mathematical structures. A class of variational schemes 
constructed for the hydrodynamic-electrodynamic model of lossless free-electron gas in a quasi-neutral background show 
good properties in secular simulations of surface plasmon polaritons [Q. Chen \emph{et al.}, Phys. Rev. E 99, 023313]. 
We show the discrete local conservation laws admitted by these schemes. Based on the gauge symmetry of the discrete action 
functional, a discrete charge conservation law is realized locally, which is consistent with the discrete Euler-Lagrange 
equations obtained from the variational schemes. Based on the discrete Euler-Lagrange equations, discrete local momentum 
and energy conservation laws are derived directly, which are rigorous in the theory. The preservation of the discrete 
local conservation laws and Lagrangian symplectic structure ensure that the numerical scheme is correct in physics.
\end{abstract}


\maketitle

\section{Introduction\label{sec:1}}

As a core concept of modern theoretical physics, Noether's theorem states that every differentiable symmetry of the 
action of a physical system has a corresponding local conservation law, which was first proven by E. Noether in 1915 \cite{Noether}. 
It is well known that the gauge symmetry gives the law of charge conservation, the spatial rotation symmetry corresponds 
to the angular momentum conservation, the spatial and temporal translation symmetries correspond to the energy and 
momentum conservations respectively \cite{Noether,Peskin}. In this perspective, physics is a kind of nature science that 
study symmetries and conservation laws. As the first principle of physics, the Hamilton's principle gives rise to physical 
laws, or in other word, the action functional determined by the path integral of the Lagrangian describes the physics of 
a given system \cite{Noether,Peskin}. The relation between symmetry of an action and conservation law of a physical system 
stated by Noether's theorem plays an important role in modern physics, especially in the gauge theory based standard 
model \cite{Peskin,Montesinos}. 

In the last half century, impressive development and significant advance of the computational physics can be observed 
in different branches of physics and their interdisciplinary fields \cite{Yee,Harrington,Taflove1,Taflove2,Anderson}. 
The numerical methods expand the theoretical studies from the linear, perturbed, and simple systems to the nonlinear, 
unperturbed, and complex systems, which bring us abundant new physics and introduce some important open problems \cite{Yee,
Harrington,Taflove1,Taflove2,Anderson}. Although the computational physics have achieved great success and become an 
indispensable part of the physical science, there are many problems in current numerical methods. The breakdown of 
conservation laws, loss of inherent mathematical structures, and import of nonphysical factors, which are introduced by 
the numerical schemes may bring pseudo physics \cite{Feng2,Hairer}. What is a correct numerical scheme in physics? The 
preservation of discrete local conservation laws and mathematical structures that are discrete analogs of the continuous 
ones admitted by the physical system, can be recognized as a standard \cite{Hairer,Qin10}. In this work, we give a detailed 
discussion about the previously constructed variational schemes for the hydrodynamic-electrodynamic model of lossless 
free-electron gas in a quasi-neutral background which are used for simulating surface plasmon polaritons (SPPs) \cite{QChen}. 
The discrete local conservation laws admitted by the variational schemes are derived, in addition to the preservation of 
Lagrangian symplectic structure, both of which ensure the good properties of the schemes in secular simulations.

Starting from the Lagrangian density of lossless free-electron gas in a quasi-neutral background with a self-consistent 
electromagnetic field given in Ref.~\cite{QChen},
\begin{eqnarray}
\mathcal{L}&=&\frac{1}{2}mn\bm{v}^2+en\bm{v}\cdot\bm{A}-e\left(n-n_{0}\right)\phi+\frac{\epsilon_{0}}{2}\left(-\bigtriangledown\phi-\frac{\partial}{\partial{t}}\bm{A}\right)^2-\frac{1}{2\mu_{0}}(\bigtriangledown\times\bm{A})^2\nonumber\\
& &+\alpha\left[\frac{\partial}{\partial{t}}n+\bigtriangledown\cdot\left(n\bm{v}\right)\right]-\lambda\left(\frac{\partial}{\partial{t}}\mu+\bm{v}\cdot\bigtriangledown\mu\right).\label{eq:1}
\end{eqnarray}
Where $n$ is the electron density, $n_{0}$ is the background particle density, $\bm{v}$ is the electron gas velocity, 
$e$ is the electron charge and $m$ is the electron mass. The electromagnetic field components $(\phi, \bm{A})$ are 
defined in an arbitary gauge. The scalar fields $\alpha$ and $\lambda$ are Lagrangian multipliers, which are introduced 
to describe the constraint equations, for the hydrodynamic system is a degenerate infinite dimensional noncanonical 
Hamiltonian system \cite{Morrison2}. The Lin's constraint factor $\mu$ recognized as a general Lagrangian coodinate is 
used to complete the vorticity structure \cite{Newcomb,Sahraoui}. By taking the total variation of the action functional 
$S=\int_{T}\int_{\Omega}\mathcal{L}{\rm{d}}x^3{\rm{d}}t$ with fixed boundary and leading the variational derivatives of 
$S$ with respect to the fields $q=(n,\bm{v},\bm{A},\phi,\alpha,\lambda,\mu)$ equal to 0, we obtain the Euler-Lagrange 
equations of lossless free-electron gas as \cite{QChen},
\begin{eqnarray}
\frac{\partial}{\partial{t}}\alpha=\frac{1}{2}m\bm{v}^{2}-\bm{v}\cdot\bigtriangledown\alpha+e\bm{v}\cdot\bm{A}-e\phi,\label{eq:2}\\
mn\bm{v}+en\bm{A}=n\bigtriangledown\alpha+\lambda\bigtriangledown\mu,\label{eq:3}\\
\frac{\partial^{2}}{\partial{t}^{2}}\bm{A}=-\frac{1}{\epsilon_{0}\mu_{0}}\bigtriangledown\times\bigtriangledown\times\bm{A}-\bigtriangledown\frac{\partial}{\partial{t}}\phi+\frac{e}{\epsilon_{0}}n\bm{v},\label{eq:4}\\
\bigtriangledown^{2}\phi+\bigtriangledown\cdot\frac{\partial}{\partial{t}}\bm{A}=-\frac{e}{\epsilon_{0}}\left(n-n_{0}\right),\label{eq:5}\\
\frac{\partial}{\partial{t}}n=-\bigtriangledown\cdot\left(n\bm{v}\right),\label{eq:6}\\
\frac{\partial}{\partial{t}}\mu=-\bm{v}\cdot\bigtriangledown\mu,\label{eq:7}\\
\frac{\partial}{\partial{t}}\lambda=-\bigtriangledown\cdot\left(\lambda\bm{v}\right).\label{eq:8}
\end{eqnarray}
Eqs.~\eqref{eq:2} and \eqref{eq:8} are Lagrangian multiplier equations. Eq.~\eqref{eq:3} defines the canonical 
momentum density of fre-electron gas. Eqs.~\eqref{eq:4} and \eqref{eq:5} are Maxwell's equations. Eq.~\eqref{eq:6} 
is the continuity equation of free-electron gas. Eq.~\eqref{eq:7} defines the dynamics of Lin's constraint field. It 
can be proven that the Euler-Lagrange equations~\eqref{eq:2}-\eqref{eq:8} equal to the well-known hydrodynamic-Maxwell 
equations \cite{QChen},
\begin{eqnarray}
\frac{\partial}{\partial{t}}n+\bigtriangledown\cdot\left(n\bm{v}\right)=0,\label{eq:9}\\
\frac{\partial}{\partial{t}}\bm{v}+\bm{v}\cdot\bigtriangledown\bm{v}=\frac{e}{m}\left(\bm{E}+\mu_{0}\bm{v}\times\bm{H}\right),\label{eq:10}\\
\frac{1}{c^{2}}\frac{\partial^{2}}{\partial{t}^{2}}\bm{A}+\bigtriangledown\times\bigtriangledown\times\bm{A}=\mu_{0}en\bm{v},\label{eq:11}\\
\bigtriangledown\cdot\frac{\partial}{\partial{t}}\bm{A}=-\frac{e}{\epsilon_{0}}\left(n-n_{0}\right).\label{eq:12}
\end{eqnarray}
Where the electromagnetic field components are defined as $\bm{E}=-\partial\bm{A}/\partial{t}$, 
$\mu_{0}\bm{H}=\bigtriangledown\times\bm{A}$, and the temporal gauge $\phi=0$ is adopted explicitly.

The hydrodynamic-electrodynamic model admits several symmetries and conservation laws. Taking the gauge transformation 
of Eq.~\eqref{eq:1} and keeping the action functional invariant, we obtain,
\begin{eqnarray}
\bm{A}^{'}&=&\bm{A}+\bigtriangledown\psi,\label{eq:13}\\
\phi^{'}&=&\phi-\frac{\partial}{\partial{t}}\psi,\label{eq:14}\\
S^{'}&=&S-\int_{T}\int_{\Omega}\left[\frac{\partial}{\partial{t}}\left(en\right)+\bigtriangledown\cdot\left(en\bm{v}\right)\right]\psi{\rm{d}}x^3{\rm{d}}t\nonumber\\
& &+\int_{T}\int_{\Omega}\left\{\frac{\partial}{\partial{t}}\left[e\left(n-n_{0}\right)\psi\right]+\bigtriangledown\cdot\left(en\psi\bm{v}\right)\right\}{\rm{d}}x^3{\rm{d}}t\nonumber\\
&=&S.\label{eq:15}
\end{eqnarray}
The third term of the expansion in Eq.~\eqref{eq:15} is a boundary integral which can be removed in the action functional. 
Because the gauge function $\psi$ is an arbitrary scalar field, a charge conservation law is obtained locally,
\begin{eqnarray}
\frac{\partial}{\partial{t}}\left(en\right)+\bigtriangledown\cdot\left(en\bm{v}\right)=0.\label{eq:16}
\end{eqnarray}
The local momentum conservation law can be derived directly from the Euler-Lagrange equations~\eqref{eq:2}-\eqref{eq:8} 
or hydrodynamic-Maxwell equations~\eqref{eq:9}-\eqref{eq:12}, for hydrodynamic equations describe the transport of 
electron gas. Multiplying Eq.~\eqref{eq:10} by $n$ and then substituting Eq.~\eqref{eq:9} into Eq.~\eqref{eq:10}, we 
obtain,
\begin{eqnarray}
\frac{\partial}{\partial{t}}\left(mn\bm{v}\right)+\bigtriangledown\cdot\left(mn\bm{v}\bm{v}\right)=en\left(\bm{E}+\mu_{0}\bm{v}\times\bm{H}\right).\label{eq:17}
\end{eqnarray}
By replacing the multiplicator $n$ with $n\bm{v}$, the local energy conservation law can be derived via the same 
approach as,
\begin{eqnarray}
\frac{\partial}{\partial{t}}\left(\frac{1}{2}mn\bm{v}^{2}\right)+\bigtriangledown\cdot\left(\frac{1}{2}mn\bm{v}^{2}\bm{v}\right)=en\bm{v}\cdot\bm{E}.\label{eq:18}
\end{eqnarray}
These local conservation laws correspond with relevant symmetries in the hydrodynamic-electrodynamic model, which should 
be preserved in the numerical schemes.

In this work we prove that the variational schemes constructed in Ref.~\cite{QChen} admit discrete local conservation 
laws. In Sec.\ref{sec:2}, the discrete local charge conservation law is proven via the gauge symmetry of the diacrete 
action functional, and the discrete Noether's theorem is obtained. In Sec.\ref{sec:3}, the discrete local momentum and 
energy conservation laws are derived directly from the discrete Euler-Lagrange equations. These desirable features and 
the conservation of Lagrangian symplectic structure proven in Ref.~\cite{QChen} make the algorithm a powerful tool in 
the study of SPPs using the hydrodynamic-electrodynamic model.

\section{Discrete Gauge Symmetry and Local Charge Conservation Law\label{sec:2}}

For the hydrodynamic and Maxwell's equations, many numerical schemes, such as the Finite-Difference Time-Domain (FDTD) 
method, Finite Volume (FV) method, Finite Element (FE) method, and Spectrum Method (SM) etc., have been developed, 
which are widely used in modern Computational Electrodynamics (CED) and Computational Fluid Dynamics (CFD) 
\cite{Yee,Harrington,Taflove1,Taflove2,Anderson,QChen0}. In order to preserve the Lagrangian symplectic structure of the 
model, we constructed a class of variational schemes in Ref.~\cite{QChen}, which are known as the structure-preserving 
geometric algorithms. The discrete Hamilton's principle based variational integrator developed by J. Marsden and M. West 
for Lagrangian systems provide us with an alternative numerical approach to achieve the preservation of discrete 
symplectic 2-form which was first realized by K. Feng in canonical Hamiltonian systems \cite{Feng2,Hairer,Marsden,Lew,
West,Qin2,Qin4,Qin5,Qin6,Qin7,Shadwick,Qin8,Qin9,Feng1,Benettin,YWu,SChin,Qin1,Morrison1,QChen2}. To apply the 
variational principle in numerical simulation, the Discrete Exterior Calculus (DEC) is introduced to discretize the 
space-time manifold and differential forms \cite{Hirani,Hiptmair,Arnold1,Arnold2,Holst}. The DEC based discrete Lagrangian 
density in Ref.~\cite{QChen} is given as,
\begin{eqnarray}
\mathcal{L}^{n{\sim}n+1}_{i,j,k}=\mathcal{L}^{n{\sim}n+1}_{{\rm{EG}}i,j,k}+\mathcal{L}^{n{\sim}n+1}_{{\rm{EM}}i,j,k}+\mathcal{L}^{n{\sim}n+1}_{{\rm{Int}}i,j,k},\label{eq:19}
\end{eqnarray}
where the subscripts EG, EM and Int indicate the electron gas, gauge field and interaction parts of the Lagrangian 
density respectively. The superscript $n$ and subscripts $i,j,k$ traverse all space-time lattice points.
\begin{eqnarray}
\mathcal{L}^{n{\sim}n+1}_{{\rm{EG}}i,j,k}&=&\frac{1}{2}mn^{n}_{i+\frac{1}{2},j+\frac{1}{2},k+\frac{1}{2}}\left(v^{n^2}_{xi+\frac{1}{2},j,k}+v^{n^2}_{yi,j+\frac{1}{2},k}+v^{n^2}_{zi,j,k+\frac{1}{2}}\right)\nonumber\\
& &+\alpha^{n+\frac{1}{2}}_{i,j,k}\left[{\rm{D}}_{t}n^{n}_{i+\frac{1}{2},j+\frac{1}{2},k+\frac{1}{2}}+\bigtriangledown_{d}\cdot\left(n\bm{v}\right)^{n+1}_{i,j,k}\right]\nonumber\\
& &-\lambda^{n}_{i+\frac{1}{2},j+\frac{1}{2},k+\frac{1}{2}}\left({\rm{D}}_{t}\mu^{n-\frac{1}{2}}_{i,j,k}+v^{n}_{xi+\frac{1}{2},j,k}{\rm{D}}_{x}\mu^{n-\frac{1}{2}}_{i,j,k}+v^{n}_{yi,j+\frac{1}{2},k}{\rm{D}}_{y}\mu^{n-\frac{1}{2}}_{i,j,k}+v^{n}_{zi,j,k+\frac{1}{2}}{\rm{D}}_{z}\mu^{n-\frac{1}{2}}_{i,j,k}\right),\label{eq:20}
\end{eqnarray}
\begin{eqnarray}
\mathcal{L}^{n{\sim}n+1}_{{\rm{EM}}i,j,k}&=&\frac{\epsilon_{0}}{2}\left[\left({\rm{D}}_{x}\phi^{n+\frac{1}{2}}_{i,j,k}+{\rm{D}}_{t}A^{n}_{xi+\frac{1}{2},j,k}\right)^{2}+\left({\rm{D}}_{y}\phi^{n+\frac{1}{2}}_{i,j,k}+{\rm{D}}_{t}A^{n}_{yi,j+\frac{1}{2},k}\right)^{2}\right.\nonumber\\
& &\left.+\left({\rm{D}}_{z}\phi^{n+\frac{1}{2}}_{i,j,k}+{\rm{D}}_{t}A^{n}_{zi,j,k+\frac{1}{2}}\right)^{2}\right]-\frac{1}{2\mu_{0}}\left(\bigtriangledown_{d}\times\bm{A}^{n}_{i,j,k}\right)^{2},\label{eq:21}
\end{eqnarray}
\begin{eqnarray}
\mathcal{L}^{n{\sim}n+1}_{{\rm{Int}}i,j,k}&=&en^{n}_{i+\frac{1}{2},j+\frac{1}{2},k+\frac{1}{2}}\left(v^{n}_{xi+\frac{1}{2},j,k}A^{n}_{xi+\frac{1}{2},j,k}+v^{n}_{yi,j+\frac{1}{2},k}A^{n}_{yi,j+\frac{1}{2},k}\right.\nonumber\\
& &\left.+v^{n}_{zi,j,k+\frac{1}{2}}A^{n}_{zi,j,k+\frac{1}{2}}\right)-e\left(n^{n}_{i+\frac{1}{2},j+\frac{1}{2},k+\frac{1}{2}}-n_{0}\right)\phi^{n+\frac{1}{2}}_{i,j,k},\label{eq:22}
\end{eqnarray}
where the temporal derivative, gradient, curl, and divergence operators in DEC framework are defined as,
\begin{eqnarray}
{\rm{D}}_{t}n^{n}_{i+\frac{1}{2},j+\frac{1}{2},k+\frac{1}{2}}=\frac{n^{n+1}_{i+\frac{1}{2},j+\frac{1}{2},k+\frac{1}{2}}-n^{n}_{i+\frac{1}{2},j+\frac{1}{2},k+\frac{1}{2}}}{\Delta{t}},\label{eq:23}
\end{eqnarray}
\begin{eqnarray}
\bigtriangledown_{d}\phi^{n+\frac{1}{2}}_{i,j,k}=\left(
\begin{array}{c}
{\rm{D}}_{x}\phi^{n+\frac{1}{2}}_{i,j,k}\\
{\rm{D}}_{y}\phi^{n+\frac{1}{2}}_{i,j,k}\\
{\rm{D}}_{z}\phi^{n+\frac{1}{2}}_{i,j,k}
\end{array}
\right)=\left(
\begin{array}{c}
\frac{\phi^{n+\frac{1}{2}}_{i+1,j,k}-\phi^{n+\frac{1}{2}}_{i,j,k}}{\Delta{x}}\\
\frac{\phi^{n+\frac{1}{2}}_{i,j+1,k}-\phi^{n+\frac{1}{2}}_{i,j,k}}{\Delta{y}}\\
\frac{\phi^{n+\frac{1}{2}}_{i,j,k+1}-\phi^{n+\frac{1}{2}}_{i,j,k}}{\Delta{z}}
\end{array}
\right),\label{eq:24}
\end{eqnarray}
\begin{eqnarray}
\bigtriangledown_{d}\times\bm{A}^{n}_{i,j,k}=\left(
\begin{array}{c}
\frac{A^{n}_{zi,j+1,k+\frac{1}{2}}-A^{n}_{zi,j,k+\frac{1}{2}}}{\Delta{y}}-\frac{A^{n}_{yi,j+\frac{1}{2},k+1}-A^{n}_{yi,j+\frac{1}{2},k}}{\Delta{z}}\\
\frac{A^{n}_{xi+\frac{1}{2},j,k+1}-A^{n}_{xi+\frac{1}{2},j,k}}{\Delta{z}}-\frac{A^{n}_{zi+1,j,k+\frac{1}{2}}-A^{n}_{zi,j,k+\frac{1}{2}}}{\Delta{x}}\\
\frac{A^{n}_{yi+1,j+\frac{1}{2},k}-A^{n}_{yi,j+\frac{1}{2},k}}{\Delta{x}}-\frac{A^{n}_{xi+\frac{1}{2},j+1,k}-A^{n}_{xi+\frac{1}{2},j,k}}{\Delta{y}}
\end{array}
\right),\label{eq:25}
\end{eqnarray}
\begin{eqnarray}
\bigtriangledown_{d}\cdot(n\bm{v})^{n}_{i,j,k}&=&\frac{n^{n}_{i+\frac{1}{2},j+\frac{1}{2},k+\frac{1}{2}}v^{n}_{xi+\frac{1}{2},j,k}-n^{n}_{i-\frac{1}{2},j+\frac{1}{2},k+\frac{1}{2}}v^{n}_{xi-\frac{1}{2},j,k}}{\Delta{x}}\nonumber\\
& &+\frac{n^{n}_{i+\frac{1}{2},j+\frac{1}{2},k+\frac{1}{2}}v^{n}_{yi,j+\frac{1}{2},k}-n^{n}_{i+\frac{1}{2},j-\frac{1}{2},k+\frac{1}{2}}v^{n}_{yi,j-\frac{1}{2},k}}{\Delta{y}}\nonumber\\
& &+\frac{n^{n}_{i+\frac{1}{2},j+\frac{1}{2},k+\frac{1}{2}}v^{n}_{zi,j,k+\frac{1}{2}}-n^{n}_{i+\frac{1}{2},j+\frac{1}{2},k-\frac{1}{2}}v^{n}_{zi,j,k-\frac{1}{2}}}{\Delta{z}}.\label{eq:26}
\end{eqnarray}

Based on these definitions, the discrete gauge transformation can be given as,
\begin{eqnarray}
\phi^{'n+\frac{1}{2}}_{i,j,k}=\phi^{n+\frac{1}{2}}_{i,j,k}-{\rm{D}}_{t}\psi^{n}_{i,j,k},\label{eq:27}\\
A^{'n}_{xi+\frac{1}{2},j,k}=A^{n}_{xi+\frac{1}{2},j,k}+{\rm{D}}_{x}\psi^{n}_{i,j,k},\label{eq:28}\\
A^{'n}_{yi,j+\frac{1}{2},k}=A^{n}_{yi,j+\frac{1}{2},k}+{\rm{D}}_{y}\psi^{n}_{i,j,k},\label{eq:29}\\
A^{'n}_{zi,j,k+\frac{1}{2}}=A^{n}_{zi,j,k+\frac{1}{2}}+{\rm{D}}_{z}\psi^{n}_{i,j,k}.\label{eq:30}
\end{eqnarray}
Substituting Eqs.~\eqref{eq:27}-\eqref{eq:30} into Eq.~\eqref{eq:19}, we obtain $\mathcal{L}^{'n{\sim}n+1}_{{\rm{EG}}i,j,k}=\mathcal{L}^{n{\sim}n+1}_{{\rm{EG}}i,j,k}$, for $\mathcal{L}^{n{\sim}n+1}_{{\rm{EG}}i,j,k}$ is guage field independent.
\begin{eqnarray}
\mathcal{L}^{'n{\sim}n+1}_{{\rm{EM}}i,j,k}&=&\frac{\epsilon_{0}}{2}\left[\left({\rm{D}}_{x}\phi^{n+\frac{1}{2}}_{i,j,k}-{\rm{D}}_{t}{\rm{D}}_{x}\psi^{n}_{i,j,k}+{\rm{D}}_{t}A^{n}_{xi+\frac{1}{2},j,k}+{\rm{D}}_{t}{\rm{D}}_{x}\psi^{n}_{i,j,k}\right)^{2}\right.\nonumber\\
& &+\left({\rm{D}}_{y}\phi^{n+\frac{1}{2}}_{i,j,k}-{\rm{D}}_{t}{\rm{D}}_{y}\psi^{n}_{i,j,k}+{\rm{D}}_{t}A^{n}_{yi,j+\frac{1}{2},k}+{\rm{D}}_{t}{\rm{D}}_{y}\psi^{n}_{i,j,k}\right)^{2}\nonumber\\
& &\left.+\left({\rm{D}}_{z}\phi^{n+\frac{1}{2}}_{i,j,k}-{\rm{D}}_{t}{\rm{D}}_{z}\psi^{n}_{i,j,k}+{\rm{D}}_{t}A^{n}_{zi,j,k+\frac{1}{2}}+{\rm{D}}_{t}{\rm{D}}_{z}\psi^{n}_{i,j,k}\right)^{2}\right]\nonumber\\
& &-\frac{1}{2\mu_{0}}\left(\bigtriangledown_{d}\times\bm{A}^{n}_{i,j,k}+\bigtriangledown_{d}\times\bigtriangledown_{d}\psi^{n}_{i,j,k}\right)^{2}\nonumber\\
&=&\mathcal{L}^{n{\sim}n+1}_{{\rm{EM}}i,j,k},\label{eq:31}
\end{eqnarray}
\begin{eqnarray}
\mathcal{L}^{'n{\sim}n+1}_{{\rm{Int}}i,j,k}&=&\mathcal{L}^{n{\sim}n+1}_{{\rm{Int}}i,j,k}+en^{n}_{i+\frac{1}{2},j+\frac{1}{2},k+\frac{1}{2}}\left(v^{n}_{xi+\frac{1}{2},j,k}{\rm{D}}_{x}\psi^{n}_{i,j,k}+v^{n}_{yi,j+\frac{1}{2},k}{\rm{D}}_{y}\psi^{n}_{i,j,k}+v^{n}_{zi,j,k+\frac{1}{2}}{\rm{D}}_{z}\psi^{n}_{i,j,k}\right)\nonumber\\
& &+e\left(n^{n}_{i+\frac{1}{2},j+\frac{1}{2},k+\frac{1}{2}}-n_{0}\right){\rm{D}}_{t}\psi^{n}_{i,j,k}\nonumber\\
&=&\mathcal{L}^{n{\sim}n+1}_{{\rm{Int}}i,j,k}-\psi^{n+1}_{i,j,k}{\rm{D}}_{t}\left(en_{i+\frac{1}{2},j+\frac{1}{2},k+\frac{1}{2}}\right)^{n}-\psi^{n}_{i,j,k}\bigtriangledown_{d}\cdot\left(en\bm{v}\right)^{n}_{i,j,k}\nonumber\\
& &+{\rm{D}}_{t}\left[e\left(n_{i+\frac{1}{2},j+\frac{1}{2},k+\frac{1}{2}}-n_{0}\right)\psi_{i,j,k}\right]^{n}+\bigtriangledown_{d}\cdot\left(en\bm{v}\zeta\psi\right)^{n}_{i,j,k}.\label{eq:32}
\end{eqnarray}
Where $\zeta$ is a spatial translation operator which can reverse the difference direction of the discrete derivative 
operators (The detailed operations of $\zeta$ can be find in App.\ref{sec:app}).

By summing over the discrete Lagrangian density on universal discrete space-time manifold, the discrete action 
functional after gauge transformation is given as,
\begin{eqnarray}
S^{'}&=&\sum_{n,i,j,k}\mathcal{L}^{'n{\sim}n+1}_{i,j,k}\nonumber\\
&=&S-\sum_{n,i,j,k}\left[{\rm{D}}_{t}\left(en_{i+\frac{1}{2},j+\frac{1}{2},k+\frac{1}{2}}\right)^{n}+\bigtriangledown_{d}\cdot\left(en\bm{v}\right)^{n+1}_{i,j,k}\right]\psi^{n+1}_{i,j,k}\nonumber\\
& &+\sum_{n,i,j,k}\left\{{\rm{D}}_{t}\left[e\left(n_{i+\frac{1}{2},j+\frac{1}{2},k+\frac{1}{2}}-n_{0}\right)\psi_{i,j,k}\right]^{n}+\bigtriangledown_{d}\cdot\left(en\bm{v}\zeta\psi\right)^{n}_{i,j,k}\right\}.\label{eq:33}
\end{eqnarray}
The third term of expansion in Eq.~\eqref{eq:33} is a discrete boundary integral. Assuming the discrete space-time 
manifold is a finite 4-dimensional torus which means the system is closed, we find that the discrete boundary integral 
in Eq.~\eqref{eq:33} equals 0 which means the integrated scalar field is sum-free. Because the discrete gauge function 
$\psi^{n}_{i,j,k}$ is arbitrary, the gauge invariance of the discrete action functional $S^{'}=S$ gives rise to,
\begin{eqnarray}
{\rm{D}}_{t}\left(en_{i+\frac{1}{2},j+\frac{1}{2},k+\frac{1}{2}}\right)^{n}+\bigtriangledown_{d}\cdot\left(en\bm{v}\right)^{n+1}_{i,j,k}=0,\label{eq:34}
\end{eqnarray}
which is the rigorous discrete local charge conservation law. Comparing Eq.~\eqref{eq:34} with the discrete equation 
of continuity obtained by the variational schemes (Eq.~(52) in Ref.~\cite{QChen}), we find that they only differ by 
a global charge factor $e$, which means that the discrete local charge conservation law \eqref{eq:34} derived from 
the discrete gauge symmetry is consistent with the discrete Euler-Lagrange equations obtained by the variational 
schemes. The previous derivation is a discrete analog of Noether's theorem in real differentiable physical system, 
which can be recognized as discrete Noether's theorem.

\section{Discrete Local Momentum and Energy Conservation Laws\label{sec:3}}

The local fields are space-time translation invariants. As a result, the Lagrangian density \eqref{eq:1} has translation 
symmetries, for it is independent with explicit coordinate variables on space-time manifold. Based on Noether's theorem, 
it is known that the hydrodynamic-electrodynamic model admits the general momentum-energy conservation laws. As a 
discrete analog of Eq.~\eqref{eq:1}, the discrete Lagrangian density \eqref{eq:19} also has space-time translation 
symmetries, for the DEC based discretization only transforms the local fields into a finite dimensional dynamical system, 
where the explicit coordinate labels are not introduced. Based on discrete Noether's theorem, the discrete Lagrangian 
momentum maps $J^{+}_{L_{d}}(q^{n},q^{n+1})$ and $J^{-}_{L_{d}}(q^{n},q^{n+1})$ in one interval are equivariant. 
Additionally, the variational principle based discrete Euler-Lagrange equations preserve the discrete Lagrangian 
momentum maps $J^{+}_{L_{d}}(q^{n-1},q^{n})=J^{-}_{L_{d}}(q^{n},q^{n+1})$ \cite{Marsden,Hairer}. As a result, the 
discrete Lagrangian momentum maps are conserved, which means that the translation invariance of Eq.~\eqref{eq:19} leads 
to discrete general momentum-energy conservation laws. The Lagrangian density constructed in constraint form brings 
extra difficult in studying symmetries. To avoid discussing the redundant symmetries of nonphysical constraint fields, 
we directly derive the local momentum and energy conservation laws from the discrete Euler-Lagrange equations.

The local momentum and energy conservation laws which are admitted by the discrete dynamical system obtained by the 
variational schemes can be directly derived from the discrete Euler-Lagrange equations (Eqs.~(44)-(54) in 
Ref.~\cite{QChen}). By taking the discrete temporal derivative ${\rm{D}}_{t}$ of the discrete canonical momentum 
density equations (Eqs.~(45)-(47) in Ref.~\cite{QChen}) and rewriting them into the compact form, we obtain,
\begin{eqnarray}
{\rm{D}}_{t}\left(mn_{i+\frac{1}{2},j+\frac{1}{2},k+\frac{1}{2}}\bm{v}_{i,j,k}\right)^{n}&=&-{\rm{D}}_{t}\left(en_{i+\frac{1}{2},j+\frac{1}{2},k+\frac{1}{2}}\bm{A}_{i,j,k}\right)^{n}+{\rm{D}}_{t}\left(n^{n}_{i+\frac{1}{2},j+\frac{1}{2},k+\frac{1}{2}}\bigtriangledown_{d}\alpha^{n-\frac{1}{2}}_{i,j,k}\right)\nonumber\\
& &+{\rm{D}}_{t}\left(\lambda^{n}_{i+\frac{1}{2},j+\frac{1}{2},k+\frac{1}{2}}\bigtriangledown_{d}\mu^{n-\frac{1}{2}}_{i,j,k}\right)\nonumber\\
&=&-en^{n}_{i+\frac{1}{2},j+\frac{1}{2},k+\frac{1}{2}}{\rm{D}}_{t}\bm{A}^{n}_{i,j,k}-e\bm{A}^{n+1}_{i,j,k}{\rm{D}}_{t}n^{n}_{i+\frac{1}{2},j+\frac{1}{2},k+\frac{1}{2}}\nonumber\\
& &+n^{n}_{i+\frac{1}{2},j+\frac{1}{2},k+\frac{1}{2}}{\rm{D}}_{t}\bigtriangledown_{d}\alpha^{n-\frac{1}{2}}_{i,j,k}+\bigtriangledown_{d}\alpha^{n+\frac{1}{2}}_{i,j,k}{\rm{D}}_{t}n^{n}_{i+\frac{1}{2},j+\frac{1}{2},k+\frac{1}{2}}\nonumber\\
& &+\lambda^{n}_{i+\frac{1}{2},j+\frac{1}{2},k+\frac{1}{2}}{\rm{D}}_{t}\bigtriangledown_{d}\mu^{n-\frac{1}{2}}_{i,j,k}+\bigtriangledown_{d}\mu^{n+\frac{1}{2}}_{i,j,k}{\rm{D}}_{t}\lambda^{n}_{i+\frac{1}{2},j+\frac{1}{2},k+\frac{1}{2}}.\label{eq:35}
\end{eqnarray}
Where the 1-forms live along the edges of the space-time lattice, e.g. $\bm{A}_{i,j,k}$ means 
$(A_{xi+\frac{1}{2},j,k},A_{yi,j+\frac{1}{2},k},A_{zi,j,k+\frac{1}{2}})$. To simplify the derivation, the gauge 
$\lambda^{n}_{i+\frac{1}{2},j+\frac{1}{2},k+\frac{1}{2}}=n^{n}_{i+\frac{1}{2},j+\frac{1}{2},k+\frac{1}{2}}$ is adopted. 
Then substituting the discrete dynaimical equations of $\alpha$, $n$, $\mu$ and $\lambda$ (Eqs.~(44), (52)-(54) in 
Ref.~\cite{QChen}) into Eq.~\eqref{eq:35}, we obtain,
\begin{eqnarray}
{\rm{D}}_{t}\left(mn_{i+\frac{1}{2},j+\frac{1}{2},k+\frac{1}{2}}\bm{v}_{i,j,k}\right)^{n}&=&-en^{n}_{i+\frac{1}{2},j+\frac{1}{2},k+\frac{1}{2}}{\rm{D}}_{t}\bm{A}^{n}_{i,j,k}+e\bm{A}^{n+1}_{i,j,k}\bigtriangledown_{d}\cdot(n\bm{v})^{n+1}_{i,j,k}-\bigtriangledown_{d}\alpha^{n+\frac{1}{2}}_{i,j,k}\bigtriangledown_{d}\cdot(n\bm{v})^{n+1}_{i,j,k}\nonumber\\
& &-\bigtriangledown_{d}\mu^{n+\frac{1}{2}}_{i,j,k}\bigtriangledown_{d}\cdot(n\bm{v})^{n+1}_{i,j,k}-n^{n}_{i+\frac{1}{2},j+\frac{1}{2},k+\frac{1}{2}}\bigtriangledown_{d}\left(\bm{v}^{n}_{i,j,k}\cdot\bigtriangledown_{d}\mu^{n-\frac{1}{2}}_{i,j,k}\right)\nonumber\\
& &+n^{n}_{i+\frac{1}{2},j+\frac{1}{2},k+\frac{1}{2}}\bigtriangledown_{d}\left(\frac{1}{2}m\bm{v}^{n^2}_{i,j,k}+e\bm{v}^{n}_{i,j,k}\cdot\bm{A}^{n}_{i,j,k}-e\phi^{n+\frac{1}{2}}_{i,j,k}-\bm{v}^{n}_{i,j,k}\cdot\bigtriangledown_{d}\alpha^{n-\frac{1}{2}}_{i,j,k}\right)\nonumber\\
&=&en^{n}_{i+\frac{1}{2},j+\frac{1}{2},k+\frac{1}{2}}\left[-{\rm{D}}_{t}\bm{A}^{n}_{i,j,k}-\bigtriangledown_{d}\phi^{n+\frac{1}{2}}_{i,j,k}+\frac{1}{2}\left(\bm{v}^{n}_{i,j,k}+\zeta\bm{v}^{n}_{i,j,k}\right)\times\bigtriangledown_{d}\times\bm{A}^{n}_{i,j,k}\right]\nonumber\\
& &-m\bigtriangledown_{d}\cdot\left(n\bm{v}\right)^{n+1}_{i,j,k}\bm{v}^{n+1}_{i,j,k}+\frac{1}{2}n^{n}_{i+\frac{1}{2},j+\frac{1}{2},k+\frac{1}{2}}\left(\bm{v}^{n}_{i,j,k}+\zeta\bm{v}^{n}_{i,j,k}\right)\nonumber\\
& &\cdot\left(e\bigtriangledown_{d}\bm{A}^{n}_{i,j,k}-\bigtriangledown_{d}\bigtriangledown_{d}\alpha^{n-\frac{1}{2}}_{i,j,k}-\bigtriangledown_{d}\bigtriangledown_{d}\mu^{n-\frac{1}{2}}_{i,j,k}\right)\nonumber\\
& &+\frac{1}{2}n^{n}_{i+\frac{1}{2},j+\frac{1}{2},k+\frac{1}{2}}\left[m\left(\bm{v}^{n}_{i,j,k}+\zeta\bm{v}^{n}_{i,j,k}\right)+e\left(\bm{A}^{n}_{i,j,k}+\zeta\bm{A}^{n}_{i,j,k}\right)\right.\nonumber\\
& &\left.-\left(\bigtriangledown_{d}\alpha^{n-\frac{1}{2}}_{i,j,k}+\zeta\bigtriangledown_{d}\alpha^{n-\frac{1}{2}}_{i,j,k}\right)-\left(\bigtriangledown_{d}\mu^{n-\frac{1}{2}}_{i,j,k}+\zeta\bigtriangledown_{d}\mu^{n-\frac{1}{2}}_{i,j,k}\right)\right]\cdot\bigtriangledown_{d}\bm{v}^{n}_{i,j,k}\nonumber\\
& &+\frac{1}{2}n^{n}_{i+\frac{1}{2},j+\frac{1}{2},k+\frac{1}{2}}\left[m\left(\bm{v}^{n}_{i,j,k}+\zeta\bm{v}^{n}_{i,j,k}\right)+e\left(\bm{A}^{n}_{i,j,k}+\zeta\bm{A}^{n}_{i,j,k}\right)\right.\nonumber\\
& &\left.-\left(\bigtriangledown_{d}\alpha^{n-\frac{1}{2}}_{i,j,k}+\zeta\bigtriangledown_{d}\alpha^{n-\frac{1}{2}}_{i,j,k}\right)-\left(\bigtriangledown_{d}\mu^{n-\frac{1}{2}}_{i,j,k}+\zeta\bigtriangledown_{d}\mu^{n-\frac{1}{2}}_{i,j,k}\right)\right]\times\bigtriangledown_{d}\times\bm{v}^{n}_{i,j,k}.\label{eq:36}
\end{eqnarray}

By taking the discrete gradient $\bigtriangledown_{d}$ of the discrete canonical momentum density equations, we obtain,
\begin{eqnarray}
-m\bigtriangledown_{d}\bm{v}^{n}_{i,j,k}=e\bigtriangledown_{d}\bm{A}^{n}_{i,j,k}-\bigtriangledown_{d}\bigtriangledown_{d}\alpha^{n-\frac{1}{2}}_{i,j,k}-\bigtriangledown_{d}\bigtriangledown_{d}\mu^{n-\frac{1}{2}}_{i,j,k}.\label{eq:37}
\end{eqnarray}
Then substituting Eq.~\eqref{eq:37} and the discrete canonical momentum density equations into Eq.~\eqref{eq:36}, we obtain,
\begin{eqnarray}
{\rm{D}}_{t}\left(mn_{i+\frac{1}{2},j+\frac{1}{2},k+\frac{1}{2}}\bm{v}_{i,j,k}\right)^{n}&=&en^{n}_{i+\frac{1}{2},j+\frac{1}{2},k+\frac{1}{2}}\left[-{\rm{D}}_{t}\bm{A}^{n}_{i,j,k}-\bigtriangledown_{d}\phi^{n+\frac{1}{2}}_{i,j,k}+\frac{1}{2}\left(\bm{v}^{n}_{i,j,k}+\zeta\bm{v}^{n}_{i,j,k}\right)\times\bigtriangledown_{d}\times\bm{A}^{n}_{i,j,k}\right]\nonumber\\
& &-m\bigtriangledown_{d}\cdot\left(n\bm{v}\right)^{n+1}_{i,j,k}\bm{v}^{n+1}_{i,j,k}-\frac{m}{2}n^{n}_{i+\frac{1}{2},j+\frac{1}{2},k+\frac{1}{2}}\left(\bm{v}^{n}_{i,j,k}+\zeta\bm{v}^{n}_{i,j,k}\right)\cdot\bigtriangledown_{d}\bm{v}^{n}_{i,j,k},\label{eq:38}
\end{eqnarray}

Eq.~\eqref{eq:38} is a discrete analog of the local momentum conservation law \eqref{eq:17} in real physical system, 
where the first term of expansion in Eq.~\eqref{eq:38} is the discrete Lorentz force density, the second and third 
terms are recognized as a discrete divergence of the momentum current density,
\begin{eqnarray}
\bigtriangledown\cdot\left(mn\bm{v}\bm{v}\right)=m\bigtriangledown\cdot\left(n\bm{v}\right)\bm{v}+mn\bm{v}\cdot\bigtriangledown\bm{v}.\label{eq:39}
\end{eqnarray}
It is very interesting that the contributions of divergence of the momentum current density in variational scheme 
based discrete dynamical system come from two adjacent temporal layers, which means half of the momentum current 
comes from the past and half of the momentum current comes from the future. The magical causality born of the symmetric 
discretization on temporal submanifold. By reviewing the discrete Euler-Lagrange equations (Eqs.~(44)-(54) in Ref.~\cite{QChen}), 
we emphasized that the variational schemes are semi-explicit, which means the temporal directions of data flows are 
hybrid. As a result, the divergence of the momentum current density in discrete local momentum conservation law split 
into two parts which live on the adjacent temporal layers. Eq.~\eqref{eq:38} provide us with a momentum conservation 
law on the discrete space-time manifold.

The discrete local energy conservation law can be derived via the same approach. Multiplying the discrete canonical 
momentum density equations by $\cdot\bm{v}^{n}_{i,j,k}$ and then taking the discrete temporal derivative of it, we 
obtain the discrete local energy conservation law. With tedious calculation, the conservation law can be written in a 
compact form as,
\begin{eqnarray}
{\rm{D}}_{t}\left(\frac{1}{2}mn_{i+\frac{1}{2},j+\frac{1}{2},k+\frac{1}{2}}\bm{v}^{2}_{i,j,k}\right)^{n}&=&\frac{e}{2}n^{n}_{i+\frac{1}{2},j+\frac{1}{2},k+\frac{1}{2}}\left(\bm{v}^{n}_{i,j,k}+\bm{v}^{n+1}_{i,j,k}\right)\cdot\left[-{\rm{D}}_{t}\bm{A}^{n}_{i,j,k}-\bigtriangledown_{d}\phi^{n+\frac{1}{2}}_{i,j,k}\right.\nonumber\\
& &\left.+\frac{1}{2}\left(\bm{v}^{n}_{i,j,k}+\zeta\bm{v}^{n}_{i,j,k}\right)\times\bigtriangledown_{d}\times\bm{A}^{n}_{i,j,k}\right]\nonumber\\
& &-\frac{m}{2}\bigtriangledown_{d}\cdot\left(n\bm{v}\right)^{n+1}_{i,j,k}\bm{v}^{n+1^2}_{i,j,k}\nonumber\\
& &-\frac{m}{4}n^{n}_{i+\frac{1}{2},j+\frac{1}{2},k+\frac{1}{2}}\left(\bm{v}^{n}_{i,j,k}+\zeta\bm{v}^{n}_{i,j,k}\right)\cdot\bigtriangledown_{d}\bm{v}^{n}_{i,j,k}\cdot\left(\bm{v}^{n}_{i,j,k}+\bm{v}^{n+1}_{i,j,k}\right).\label{eq:40}
\end{eqnarray}
Eq.~\eqref{eq:40} is a discrete analog of the local energy conservation law \eqref{eq:18} in real physical system, 
where the first term of expansion in Eq.~\eqref{eq:40} is the discrete power density induced by Lorentz force, the 
second and third terms can be recognized as a discrete divergence of the energy current density,
\begin{eqnarray}
\bigtriangledown\cdot\left(\frac{1}{2}mn\bm{v}^{2}\bm{v}\right)=\frac{m}{2}\left[\bigtriangledown\cdot\left(n\bm{v}\right)\bm{v}+2n\bm{v}\cdot\bigtriangledown\bm{v}\right]\cdot\bm{v}.\label{eq:41}
\end{eqnarray}
Just as the discrete local momentum conservation law, the contributions of power density induced by Lorentz force come 
from two adjacent temporal layers, which correspond to relevant velocity $\bm{v}^{n}$ or $\bm{v}^{n+1}$ respectively. 
The discrete divergence of the energy current density in Eq.~\eqref{eq:40} also split into two parts, where the second 
term only comes from the future, but the third term is past-future hybrid. Different from the continuous dynamical 
system, the discrete local energy conservation law \eqref{eq:40} include a contribution of magnetic power,
\begin{eqnarray}
P_{M}=\frac{e}{4}n^{n}_{i+\frac{1}{2},j+\frac{1}{2},k+\frac{1}{2}}\left(\bm{v}^{n}_{i,j,k}+\bm{v}^{n+1}_{i,j,k}\right)\cdot\left[\left(\bm{v}^{n}_{i,j,k}+\zeta\bm{v}^{n}_{i,j,k}\right)\times\bigtriangledown_{d}\times\bm{A}^{n}_{i,j,k}\right].\label{eq:42}
\end{eqnarray}
This nonphysical term comes from the discretization, which is inherent in numerical schemes. Obviously, 
Eq.~\eqref{eq:42} admits $\lim_{\Delta{V},\Delta{t}\to0}P_{M}=0$, which means the discrete conservation law will 
returen to the real one in continuous limit. Eq.~\eqref{eq:40} provide us with a energy conservation law on the 
discrete space-time manifold.

\section{Discussion and Conclusion}

In summary, we derived the rigorous local charge, momentum and energy conservation laws of the discrete dynamical system 
obtained by the variational schemes constructed for simulating the SPPs. In the derivation of local charge conservation 
law, the gauge symmetry of discrete action functional was proven, which is recognized as a discrete analog of Noether's 
theorem in real differentiable physical system. We emphasize that the realization of discrete Noether's theorem 
is significant for a discrete dynamical system, for it means that the discrete dynamical system has intrinsic symmetries 
and local conservation laws which are essential for a real differentiable physical system. As the hydrodynamic system is 
a highly degenerate infinite dimensional noncanonical Hamiltonian system, the Lagrangian density used for constructing the 
variational schemes is in constraint form. To avoid discussing the redundant symmetries of nonphysical constraint fields, 
we directly derived the local momentum and energy conservation laws from the discrete Euler-Lagrange equations, for 
hydrodynamic equations describe the transport of electron gas. Just as the results shown in previous derivation, the 
discrete local momentum and energy conservation laws obtained are rigorous and closed in the theory. As an local 
momentum conservation law of the discrete dynamical system, we find that half of the momentum current comes from the 
past temporal layer and half of the momentum current comes from the future temporal layer. We attribute this magical 
causality to the symmetric discretization on temporal submanifold, and the hybrid data flow of semi-explicit schemes 
obtained by the discrete variational principle can be recognized as a support. The local energy conservation law obtained 
is more complicated in causality, for energy is a higher order dynamical quantity than momentum. The power or energy 
flow correspond to momentum current split into two parts, which live on the adjacent temporal layers. Different from the 
real differentiable physical system, the local energy conservation law of the discrete dynamical system include a 
contribution of magnetic power. This nonphysical term born of the discretization. By reviewing Eq.~\eqref{eq:42}, we 
find that the translations of velocity components in space-time lattice lead to weak nonorthogonality between momentum 
current and magnetic force, which introduce the weak magnetic power. Obviously, this term vanishes in continuous limit. 
The symmetries and local conservation laws of variational schemes derived are rigorous and self-consistent. Additionally, 
In Ref.~\cite{QChen}, we have proven the conservation of Lagrangian symplectic structure and the equality between 
variational schemes for gauge field components and traditional FDTD method for electromagnetic field components. In DEC 
framework, the conservation of electromagnetic field energy admitted by FDTD method is well known and fully discussed 
\cite{Chew,Qin10}, which means the discrete conservation laws both in electron gas and gauge field are realized. The 
symmetries, local conservation laws and preservation of Lagrangian symplectic structure ensure the discrete dynamical 
system obtained by variational schemes for SPPs simulations is correct in physics. The good properties of these schemes 
in secular simulations shown in Ref.~\cite{QChen} are natural consequences of the conservation discussed in this work.

\begin{acknowledgments}
Q. Chen would like to thank H. Qin and J. Xiao for the considerable help in differential geometry and field theory 
at the University of Science and Technology of China. This work was supported by the National Nature Science Foundation 
of China (NSFC-11805273), the CEMEE State Key Laboratory Foundation (CEMEE-2018Z0104B) and the Testing Technique 
Foundation for Young Scholars.
\end{acknowledgments}

\appendix
\section{The Spatial Translation Operator\label{sec:app}}
To simplify the derivation, we define a spatial translation operator $\zeta$ which can be used to realize the 
coordinate index translation of the discrete differential forms. Here we illustrate the detailed operations of 
$\zeta$ in different expansions.
\begin{eqnarray}
\bigtriangledown_{d}\cdot\left(\bm{v}\zeta\psi\right)^{n}_{i,j,k}&=&\frac{v^{n}_{xi+\frac{1}{2},j,k}\psi^{n}_{i+1,j,k}-v^{n}_{xi-\frac{1}{2},j,k}\psi^{n}_{i,j,k}}{\Delta{x}}+\frac{v^{n}_{yi,j+\frac{1}{2},k}\psi^{n}_{i,j+1,k}-v^{n}_{yi,j-\frac{1}{2},k}\psi^{n}_{i,j,k}}{\Delta{y}}\nonumber\\
& &+\frac{v^{n}_{zi,j,k+\frac{1}{2}}\psi^{n}_{i,j,k+1}-v^{n}_{zi,j,k-\frac{1}{2}}\psi^{n}_{i,j,k}}{\Delta{z}},\label{eq:a1}
\end{eqnarray}
\begin{eqnarray}
{\zeta}\bm{A}^{n}_{i,j,k}\cdot\bigtriangledown_{d}\bm{v}^{n}_{i,j,k}\nonumber\\
=\left(
\begin{array}{c}
A^{n}_{xi+\frac{3}{2},j,k}{\rm{D}}_{x}v^{n}_{xi+\frac{1}{2},j,k}+A^{n}_{yi+1,j+\frac{1}{2},k}{\rm{D}}_{y}v^{n}_{xi+\frac{1}{2},j,k}+A^{n}_{zi+1,j,k+\frac{1}{2}}{\rm{D}}_{z}v^{n}_{xi+\frac{1}{2},j,k}\\
A^{n}_{xi+\frac{1}{2},j+1,k}{\rm{D}}_{x}v^{n}_{yi,j+\frac{1}{2},k}+A^{n}_{yi,j+\frac{3}{2},k}{\rm{D}}_{y}v^{n}_{yi,j+\frac{1}{2},k}+A^{n}_{zi,j+1,k+\frac{1}{2}}{\rm{D}}_{z}v^{n}_{yi,j+\frac{1}{2},k}\\
A^{n}_{xi+\frac{1}{2},j,k+1}{\rm{D}}_{x}v^{n}_{zi,j,k+\frac{1}{2}}+A^{n}_{yi,j+\frac{1}{2},k+1}{\rm{D}}_{y}v^{n}_{zi,j,k+\frac{1}{2}}+A^{n}_{zi,j,k+\frac{3}{2}}{\rm{D}}_{z}v^{n}_{zi,j,k+\frac{1}{2}}
\end{array}
\right),\label{eq:a2}
\end{eqnarray}
\begin{eqnarray}
{\zeta}\bm{A}^{n}_{i,j,k}\times\bigtriangledown_{d}\times\bm{v}^{n}_{i,j,k}\nonumber\\
=\left[
\begin{array}{c}
A^{n}_{yi+1,j+\frac{1}{2},k}\left({\rm{D}}_{x}v^{n}_{yi,j+\frac{1}{2},k}-{\rm{D}}_{y}v^{n}_{xi+\frac{1}{2},j,k}\right)-A^{n}_{zi+1,j,k+\frac{1}{2}}\left({\rm{D}}_{z}v^{n}_{xi+\frac{1}{2},j,k}-{\rm{D}}_{x}v^{n}_{zi,j,k+\frac{1}{2}}\right)\\
A^{n}_{zi,j+1,k+\frac{1}{2}}\left({\rm{D}}_{y}v^{n}_{zi,j,k+\frac{1}{2}}-{\rm{D}}_{z}v^{n}_{yi,j+\frac{1}{2},k}\right)-A^{n}_{xi+\frac{1}{2},j+1,k}\left({\rm{D}}_{x}v^{n}_{yi,j+\frac{1}{2},k}-{\rm{D}}_{y}v^{n}_{xi+\frac{1}{2},j,k}\right)\\
A^{n}_{xi+\frac{1}{2},j,k+1}\left({\rm{D}}_{z}v^{n}_{xi+\frac{1}{2},j,k}-{\rm{D}}_{x}v^{n}_{zi,j,k+\frac{1}{2}}\right)-A^{n}_{yi,j+\frac{1}{2},k+1}\left({\rm{D}}_{y}v^{n}_{zi,j,k+\frac{1}{2}}-{\rm{D}}_{z}v^{n}_{yi,j+\frac{1}{2},k}\right)
\end{array}
\right].\label{eq:a3}
\end{eqnarray}

\nocite{*}
\providecommand{\noopsort}[1]{}\providecommand{\singleletter}[1]{#1}%

\end{document}